\documentclass[pra,amsmath,amssymb,superscriptaddress,twocolumn]{revtex4-1}

\usepackage{amsmath}
\usepackage{amssymb}
\usepackage{latexsym}
\usepackage{amsfonts}
\usepackage{epsfig}
\usepackage{psfrag}
\usepackage{graphicx}
\usepackage{hyperref}

\usepackage{color}

\definecolor{black}{rgb}{0,0,0}
\definecolor{blue}{rgb}{0,0,1}
\definecolor{green}{rgb}{0,1,0}
\definecolor{red}{rgb}{1,0,0}
\definecolor{brown}{rgb}{0.4,0.2,0}
\definecolor{darkgreen}{rgb}{0,0.7,0}

\renewcommand{\vec}[1]{\boldsymbol #1}
\newcommand{\ket}[1]{\left|#1\right>}
\newcommand{\bra}[1]{\left<#1\right|}

\newcommand{\nn}{\nonumber\\}

\newcommand{\bea}{\begin{eqnarray}}
\newcommand{\ea}{\end{eqnarray}}
\newcommand{\eea}{\end{eqnarray}}

\newcommand{\matr}[1]{\boldsymbol{#1}}
\renewcommand{\vec}[1]{\mathbf{#1}}

\begin{document}

\title{Ion trap analogue of particle creation in black holes and cosmology}

\author{Christian Fey}
\affiliation{Zentrum f\"ur optische Quantentechnologien,
Luruper Chaussee 149, 22761 Hamburg, Universit\"at Hamburg, Germany}

\author{Tobias Schaetz}
\affiliation{Albert-Ludwigs-Universit\"at Freiburg, Physikalisches Institut,
Hermann-Herder-Strasse 3, 79104 Freiburg, Germany}

\author{Ralf Sch\"utzhold}
\email[e-mail:\,]{ralf.schuetzhold@uni-due.de}
\affiliation{Fakult\"at f\"ur Physik, Universit\"at Duisburg-Essen,
Lotharstrasse 1, 47057 Duisburg, Germany}

\date{\today}

\begin{abstract}
We consider the transversal modes of ions in a linear radio
frequency (rf) trap where we control the time dependent axial
confinement to show that we can excite quanta of motion via a
two-mode squeezing process.
This effect is analogous to phenomena predicted to occur during the
evaporation of black holes and cosmological particle creation, in
general out of reach for experimental investigation.
As substantial advantage of this proposal in comparison to previous
ones we propose to exploit radial and axial modes simultaneously to
permit experimental access of these effects based on
state-of-the-art technology. In addition, we propose to create and
explore entanglement, starting with two ions, and relate the results
to fundamental aspects of the entropy of black holes.
\end{abstract}


\maketitle

\section{Introduction}

It is a fundamental prediction of quantum field theory that extreme
conditions, such as non-adiabatic dynamics, can create pairs of
particles out of the quantum vacuum.
Examples are Hawking radiation, i.e., black hole evaporation, and
cosmological particle creation \cite{birrell_quantum_1982}.
To provide an intuitive picture of such an effect, let us imagine
two pendula coupled by a spring.
The classical ground states with and without spring remain
identical, however, the ground states of the quantum version differ
in a fundamental way.
Without the spring, we describe the system by a product of the
individual ground states, while the two coupled pendula require
entanglement of the non-separable state, see \cite{reznik_2005}.
Now, we remove the spring instantaneously such that the system has
no time evolve to react, we end up with two pendula which are not in
their individual ground states, i.e., excited.
The entanglement of the state corresponds to the correlation between
the two pendula, e.g., if pendulum one was in the first excited
state, then the second one has to match the excitation -- while the
total state of the system remains a pure state..
This entanglement also implies that if we consider one pendulum
only, by tracing over the degrees of freedom of the second one, the
effective state of pendulum one will be indistinguishable from a
thermal (i.e., mixed) state.

In quantum field theory, this instantaneous or non-adiabatic removal
of the spring is predicted to be caused by extreme circumstances,
such as during the inflationary part of the expansion of the
universe or in vicinity of a black hole, when wave-packets get torn
apart.
In the latter case, the entanglement between the two ``pendula''
(one inside and the other outside the horizon) explains the thermal
character of Hawking radiation.
Here, we propose an experimentally realizable analogue of this
effects based on trapped ions.
The radial modes of the two or more ions represent the
two quantum pendula while the spring is analogues to their Coulomb
interaction within the axial trapping potential. We define the
amplitude and the evolution in time of the latter by applying
potentials to additional electrodes, controlling the axial motion of
the ions and their mutual distance, respectively.
Due to the unique control and accurate detection of the elctronic
and motional degrees of freedom, trapped ions are very good
candidates for investigating these quantum effect, see also
\cite{schutzhold_analogue_2007}.
 Further examples for the simulation
of relativistic effects in ion traps can be found in
\cite{milburn_2005,schaetz_2007,solano_2011}.
%
%

\section{Excitation of Phonons}

We investigate a system of $N$ ions of the identical species in a
harmonic trapping potential characterized by a constant radial
secular frequency $\omega^2_\text{rad}$, provided by time-averaging
the rf-potential. In axial direction, we specify the time-dependent
confinement by $\omega^2_\text{ax}(t)$.
This system is a generalization of the one-dimensional approach
treated in \cite{schutzhold_analogue_2007}, where only the motion
along the axial direction has been investigated.
Here we assume that the radial confinement is always stronger than the axial
one, i.e., $\omega^2_\text{rad}> \omega^2_\text{ax}(t)$.
The classical equation of motion of the $k$-th ion with coordinate
$\vec{r}_k$ reads then
\begin{equation}
\ddot{\vec{r}}_k+
\left(\begin{matrix}
\omega^2_\text{ax}(t)&0&0 \\
0 &\omega^2_\text{rad} &0 \\
0&0& \omega^2_\text{rad}
\end{matrix}\right)\cdot\vec{r}_k
= \gamma \sum_{l \neq k}^N \frac{\vec{r}_k-\vec{r}_l}{|\vec{r}_k-\vec{r}_l|^3}
\ ,
\label{eqn:equation_of_motion_ions}
\end{equation}
where the constant $\gamma$ encodes the strength of the Coulomb repulsion
between the ions.
In the following, we focus on solutions of
(\ref{eqn:equation_of_motion_ions}) starting in the static
equilibrium positions $\vec{r}_k(t_\text{in}):=\vec{r}_k^\text{eq}:=
(x_k^\text{eq},0,0)^T $.
The solutions of (\ref{eqn:equation_of_motion_ions})
can be written as  $\vec{r}_k(t) = b(t)\vec{r}_k^\text{eq}$,
where the scale parameter $b(t)$ fulfills
\begin{equation}
\ddot{b}(t)+\omega^2_{ax}(t) b(t)
=
\frac{\left(\omega_\text{ax}^\text{in}\right)^2 }{b^2(t)} \ .
\label{eqn:b}
\end{equation}
The boundary conditions are $b(t_\text{in})=1$ and $\dot{b}(t_\text{in})=0$.
This means the classical solution is fully determined as a time dependent
rescaling of the initial equilibrium positions.

However, the ions are quantum particles described by a wave function
of a certain width, individual measurements of their positions have
to deviate from and fluctuate around their classically predictable
positions, revealing quantum fluctuations.
Their position operator can be written as $\hat{\vec{r}}_k(t)=b(t)
\vec{r}^\text{eq} + \delta \hat{\vec{r}}_k$ and in a semiclassical
approximation, we assume that the deviations $\delta
\hat{\vec{r}}_k$ remain small (because the mass of the ions being
large corresponding to a narrow width of their ground state wave
function).

Linerarization and diagonalization of  (\ref{eqn:equation_of_motion_ions})
yields then the Heisenberg equation of motion for the normal modes (phonons).
While the axial phonons have been discussed in \cite{schutzhold_analogue_2007} we focus here on the radial phonons satisfying
\begin{equation}
\left(\frac{d^2}{dt^2}+\Omega_\kappa^2(t)\right)  \delta \hat{y}_{\kappa}= 0 \ .
\label{eqn:radial}
\end{equation}
Every radial normal mode $\delta \hat y_\kappa$  can be associated to
one individual harmonic oscillator with time dependent normal mode
frequency
\begin{equation}
\Omega_\kappa^2(t)= \omega^2_\text{rad} - \frac{\omega^2_{\kappa}}{b^3(t)}
\label{eqn:normal_mode_frequencies}
\end{equation}
where $\omega^2_{\kappa}\geq 0$ is the $\kappa$-th eigenvalue of the matrix
\begin{equation}
M_{kl}=
\delta_{kl}\sum\limits_{j \neq k}^N \frac{\gamma}{{|x_k^\text{eq}-x_j^\text{eq}|}^3}
-\frac{\gamma(1-\delta_{kl})}{{|x_k^\text{eq}-x_l^\text{eq}|}^3}
\ .
\label{eqn:M}
\end{equation}
%
%
Especially for the center of mass mode we have $\omega_0=0$ and for the
rocking mode $\omega_1=\omega_\text{ax}^\text{in}$.

In the following we show how the time-dependence of the normal mode
frequencies $\Omega_\kappa(t)$  can lead to the excitation of phonons.
At the initial instant $t_\text{in}$ we express the position operator of
each normal mode in terms of the harmonic oscillator ladder operators as
\begin{equation}
\delta \hat{y}_\kappa(t_\text{in})
=
\frac{1}{\sqrt{2 \Omega_{\kappa}(t_\text{in})}}\,\hat{a}_{\kappa}^\text{in}
+ {\rm h.c.} \ .
\label{eqn:in-solution}
\end{equation}
For another given instant $t_\text{out} > t_\text{in}$, the operator evolves
under the Heisenberg equation (\ref{eqn:radial}) into
\begin{equation}
\delta \hat{y}_\kappa(t_\text{out})
=
\frac{1}{\sqrt{2 \Omega_{\kappa}(t_\text{out})}}\, \hat{a}_{\kappa}^\text{out}
+ {\rm h.c.} \ ,
\end{equation}
where the final creation/annihilation operators
$\hat{a}_{\kappa}^{\dagger \text{out}}$/$\hat{a}_{\kappa}^{\text{out}}$
are linked to the initial ones via the Bogoliubov transformation
\begin{equation}
\hat{a}_{\kappa}^\text{out}
=
\alpha_{\kappa}^* \hat{a}_{\kappa}^\text{in} -
\beta_{\kappa}^* \hat{a}_{\kappa}^{\dagger \text{in}} \ .
\label{eqn:boboliubov_ionen}
\end{equation}
with the (complex) Bogoliubov coefficients $\alpha_\kappa$ and $\beta_\kappa$.
For the initial ground state $\ket{\Psi(t_\text{in})}=\ket{0}$ in
the $\kappa$-th radial mode, the mean number of created phonons is
given by
\begin{equation}
\left<\hat{n}_\kappa^\text{out}\right>
=
\bra{\Psi(t_\text{in})}
\hat{a}_{\kappa}^{\dagger \text{ out}} \hat{a}_{\kappa}^\text{out}
\ket{\Psi(t_\text{in})}
=
|\beta_\kappa|^2
\ .
\end{equation}
Hence, phonon creation takes place depending on the temporal
evolution of $\Omega_\kappa(t)$ from $t_\text{in}$ to $t_\text{out}$,
if $|\beta_\kappa|>0$.
Or in other words: The classical motion along the $x$-axis induces
the creation of phonons in the radial direction.

The generators of the Bogoliubov transformation (\ref{eqn:boboliubov_ionen})
are Squeezing Operators.
Therefore the time evolution of the initial ground state
$\ket{\Psi(t_\text{in})}$  is given by
\begin{align}
\ket{\Psi(t_\text{out})}
&=
\hat{\mathcal{S}}_\xi \ket{0} \exp\left\{\frac{1}{2}
\sum \limits_{\kappa} \xi_{\kappa} \left(\hat{a}_{\kappa}^{\dagger \text{in}} \right)^2 -
{\rm h.c.} \right\} \ket{0}
\nonumber \\
&=
\ket{0} + \frac{1}{\sqrt{2}} \sum_\kappa \xi_\kappa \ket{2_\kappa} +
\mathcal{O}(\xi_\kappa^2) \ .
\label{eqn:squeez_ionen_aktiv}
\end{align}
where the Squeezing parameter $\xi_\kappa$ is linked to the Bogoliubov coefficients via $|\beta_\kappa|=\sinh|\xi_\kappa|$ and
$\arg\xi_\kappa=-(\arg\alpha_\kappa + \arg\beta_\kappa)$.
Formula (\ref{eqn:squeez_ionen_aktiv}) features the characteristics
of a squeezing operation, the creation of particles (here phonons)
in pairs.

\section{Excitation models for two ions}
\label{sec:excitation_models}

In the following we focus on the case of $N=2$ ions and investigate the
phonon creation induced by different axial motions of the ions.
Firstly by a collision between the ions described by a scale functions
$b_\text{col}(t)$ and secondly by an expansion of the ions corresponding
to a scale function $b_\text{ex}(t)$.
The time dependence of the axial confinement necessary to generate a given
scale function $b(t)$ can be deduced from (\ref{eqn:b}) to
\begin{equation}
\omega_\text{ax}(t)
=
\sqrt{
\frac{\left(\omega_\text{ax}^\text{in}\right)^2}{b^3(t)}-
\frac{\ddot{b}(t)}{b(t)}
} \ .
\label{eqn:omega_col}
\end{equation}
We focus here on trajectories where $\omega_\text{ax}(t) \in \mathbb{R}$.
However, there exist also trajectories $b(t)$ that can only be realized for
temporarily negative $\omega_\text{ax}^2$, that means for temporarily repulsive
trapping potentials.

The scale function is linked to the (classical) mutual distance of
the ions via
\begin{equation}
\Delta x (t)= x_1(t) - x_2(t) =b(t)  \Delta x^\text{eq} \ .
\end{equation}
In the radial direction we have the two phonon modes
\begin{equation}
\delta\hat{y}_\pm
=
\frac{1}{\sqrt{2}} \left(\delta \hat{y}_1 \pm \delta \hat{y}_2 \right) \ .
\label{eqn:ypm}
\end{equation}
This is the center of mass mode $\delta\hat{y}_+$ with the frequency
$\Omega^2_+=\omega^2_\text{rad}$ and the rocking mode $\delta\hat{y}_-$
with the frequency
$\Omega^2_-(t)=
\omega^2_\text{rad}-\left(\omega_\text{ax}^\text{in}\right)^2 / b(t)^3$.
With (\ref{eqn:normal_mode_frequencies}) the equation of motion for the
rocking mode phonons is
\begin{equation}
\left(\frac{d^2}{dt^2}+ \Omega_-^2(t)\right)  \delta \hat{y}_{-}= 0 \ .
\label{eqn:rockingmode}
\end{equation}

\subsection{Collision model}

We consider now a special scale function
\begin{equation}
b_\text{col}(t)
=
\left(
1+ \frac{\Delta \Omega_\text{col}^2}{\left(\omega_\text{ax}^\text{in}\right)^2}
\frac{1}{\cosh^2 (\omega_\text{col} t)}
\right)^{-\frac{1}{3}}
\label{eqn:modb}
\end{equation}
that parametrizes a collision between the ions.
Starting at $t_\text{in} \to - \infty$ in the equilibrium position
with $b_\text{col}(t_\text{in})=1$, the ions approach each other,
reach for $t=0$ a minimal axial distance $\Delta x _\text{min}$ at
the turning point and finally return to their initial positions for
$t_\text{out} \to + \infty$ with $b(t_\text{out})=1$.
The parameter $\Delta \Omega_\text{col}^2$ describes the change in
the rocking mode frequency $\Omega_-^2$ from $t_\text{in}$ to $t=0$ and
determines the minimal distance of the ions.
The parameter $\omega_\text{col}$  determines the characteristic
time scale of the collision.
Eq.~(\ref{eqn:rockingmode}) can be solved for $b_\text{col}(t)$ in terms of
hypergeometric functions whose asymptotic behaviour is known for
$t \to \pm \infty$.
As shown in appendix \ref{sec:beta_col} this yields the Bogoliubov coefficient
\begin{equation}
|\beta^\text{col}_-|^2
=
\left|
\frac{\cosh \bigg(\displaystyle\frac{\pi}{2} \sqrt{\frac{4 \Delta \Omega_\text{col} ^2}{\omega_\text{col} ^2}-1}\bigg)}{\sinh \left(\displaystyle\frac{\pi \Omega_\text{in}}{\omega_\text{col} }\right)}
\right|^2 \ ,
\label{eqn:betamodell}
\end{equation}
where $\Omega_\text{in}= \Omega_-(t_\text{in})$.
Here we focus on a regime of moderate and slow collisions,
where $\omega_\text{col} \ll \Delta \Omega_\text{col} < \Omega_\text{in} $.
Especially, this implies that the system never reaches critical
points with $\Omega_-=0$, where the classical radial motion becomes
unstable and the linear chain features a phase transition into a
two-dimensional zig-zag structure \cite{walther_1992}.
Under these assumptions, Eq.~(\ref{eqn:betamodell}) can be approximated as
\begin{equation}
|\beta^\text{col}_-|^2
\approx
\exp\left[
- 2 \pi \frac{(\Omega_\text{in}-\Delta \Omega_\text{col})}{\omega_\text{col}}
\right] \ .
\label{eqn:beta_wkb}
\end{equation}
That means that particle creation becomes only important if
$\Omega_\text{in}-\Delta \Omega_\text{col}$ is chosen sufficiently
small
\begin{equation}
\Omega_\text{in}-\Delta \Omega_\text{col}
=
\mathcal{O}\left(\omega_\text{col} \right)
\label{eqn:exponent_kriterium_col}
\end{equation}
and is exponentially suppressed for
$\Omega_\text{in}-\Delta \Omega_\text{col}\gg\omega_\text{col}$.
In fact, this statement is valid for generic scale functions, as
long as the collision fulfills the given assumptions.
Performing a WKB approximation in appendix \ref{sec:WKB} we show
that the mean number of phonons is mainly dominated by the relation
between two parameters: the normal mode frequency $\Omega_-^2(0)$ and
its curvature $\frac{d^2}{dt^2} \Omega_-^2(0)$, both evaluated at
the turning point, which occurs at $t=0$.
As $\Omega_-(t)^2$ is linked to the ion
trajectories via (\ref{eqn:normal_mode_frequencies}) and
(\ref{eqn:equation_of_motion_ions}) it is equivalent to state that
the mean number of phonons after the collision is mainly dominated
by the two parameters
\begin{equation}
p_1:= \left(\frac{\Delta x^\text{eq}}{\Delta x^\text{min}}\right)^3
\end{equation}
and
\begin{equation}
p_2:= \left(
\frac{\omega_\text{ax}(t=0)}{\omega_\text{ax}^\text{in}}
\right)^2 \ ,
\end{equation}
where $\omega_\text{ax}\left(t=0\right)$ describes the axial
confinement at the instance when the ions reach their turning point.
%
%
For example a model collision with trajectory $b_\text{col}(t)$
yielding the two related values $p_1$ and $p_2$ is obtained by
choosing
\begin{equation}
\Delta \Omega_\text{col}^2 (p_1,p_2) =
\left(\omega_\text{ax}^\text{in}\right)^2 \left(p_1 -1 \right)
\label{eqn:subst1}
\end{equation}
and
\begin{equation}
\omega^2_\text{col} (p_1,p_2) =
\left(\omega_\text{ax}^\text{in}\right)^2
\frac{3 p_1  \left(p_1 -p_2 \right)}{2 (p_1 -1)}
\ .
\label{eqn:subst2}
\end{equation}
We can take advantage of this to obtain approximations for the Bogoliubov
coefficients $\beta_-$ of moderate and slow collision with trajectories
qualitatively similar to (\ref{eqn:betamodell}).
Such a collision with given parameters $p_1$ and $p_2$ will lead to similar
phonon excitations as model collisions (\ref{eqn:betamodell}) having
identical parameters.
Therefore the Bogoliubov coefficient $|\beta_-|^2$ can be approximated as
\begin{equation}
|\beta_-|^2 \approx |\beta^\text{col}_-(p_1,p_2)|^2 \ ,
\label{eqn:betap1p2}
\end{equation}
where $\beta^\text{col}_-(p_1,p_2)$ denotes $\beta_-^\text{col}$ from
(\ref{eqn:betamodell}) with the substitutions (\ref{eqn:subst1}) and
(\ref{eqn:subst2}).

\begin{figure}
\includegraphics[width=0.9\linewidth]{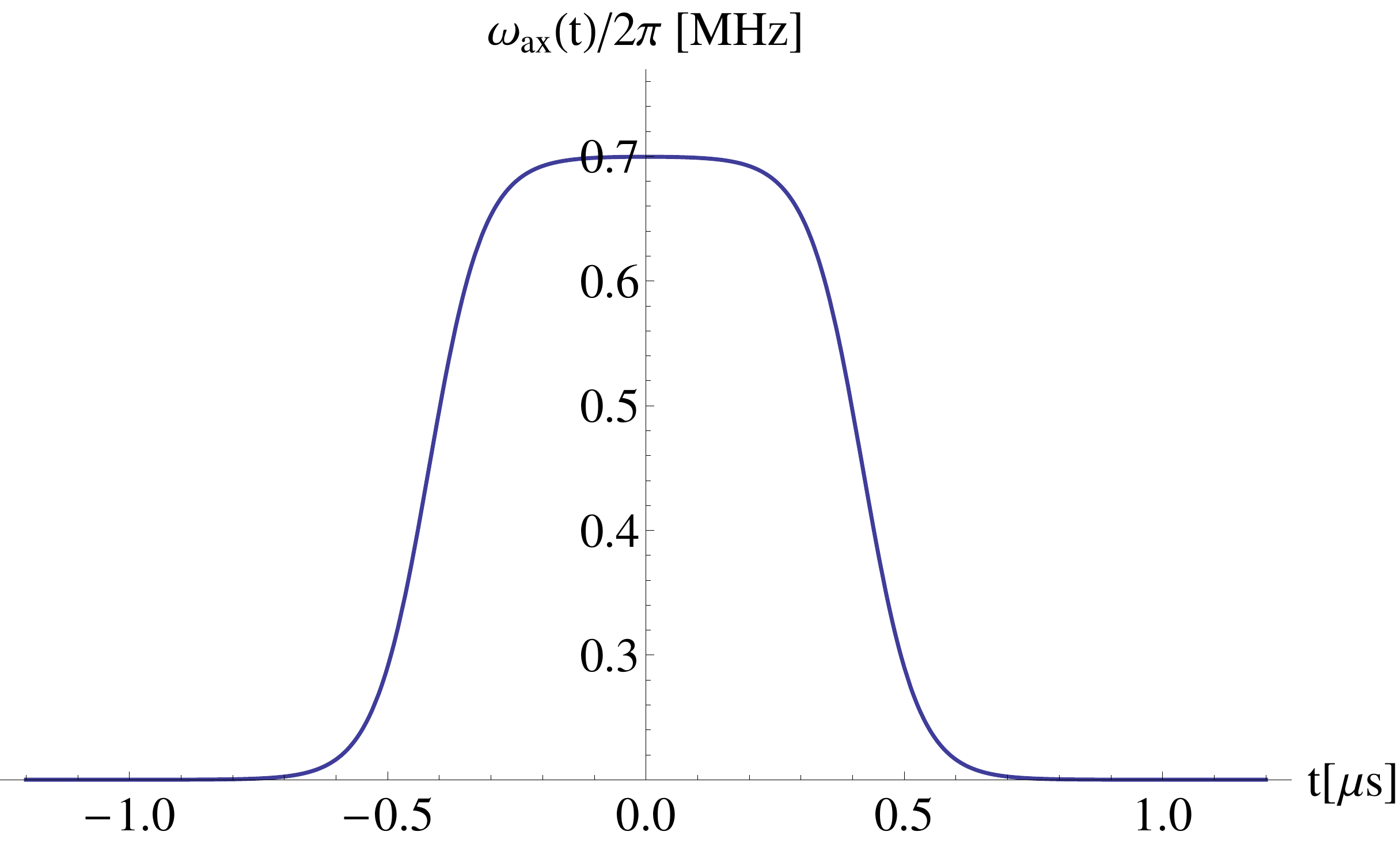}
\caption{Example of a time dependent axial confinement characterized by $\omega_\text{ax}(t)$
that leads to the classical motion illustrated in Fig. \ref{fig:delta_x}.}
\label{fig:wax}
\end{figure}

\begin{figure}
\includegraphics[width=0.9\linewidth]{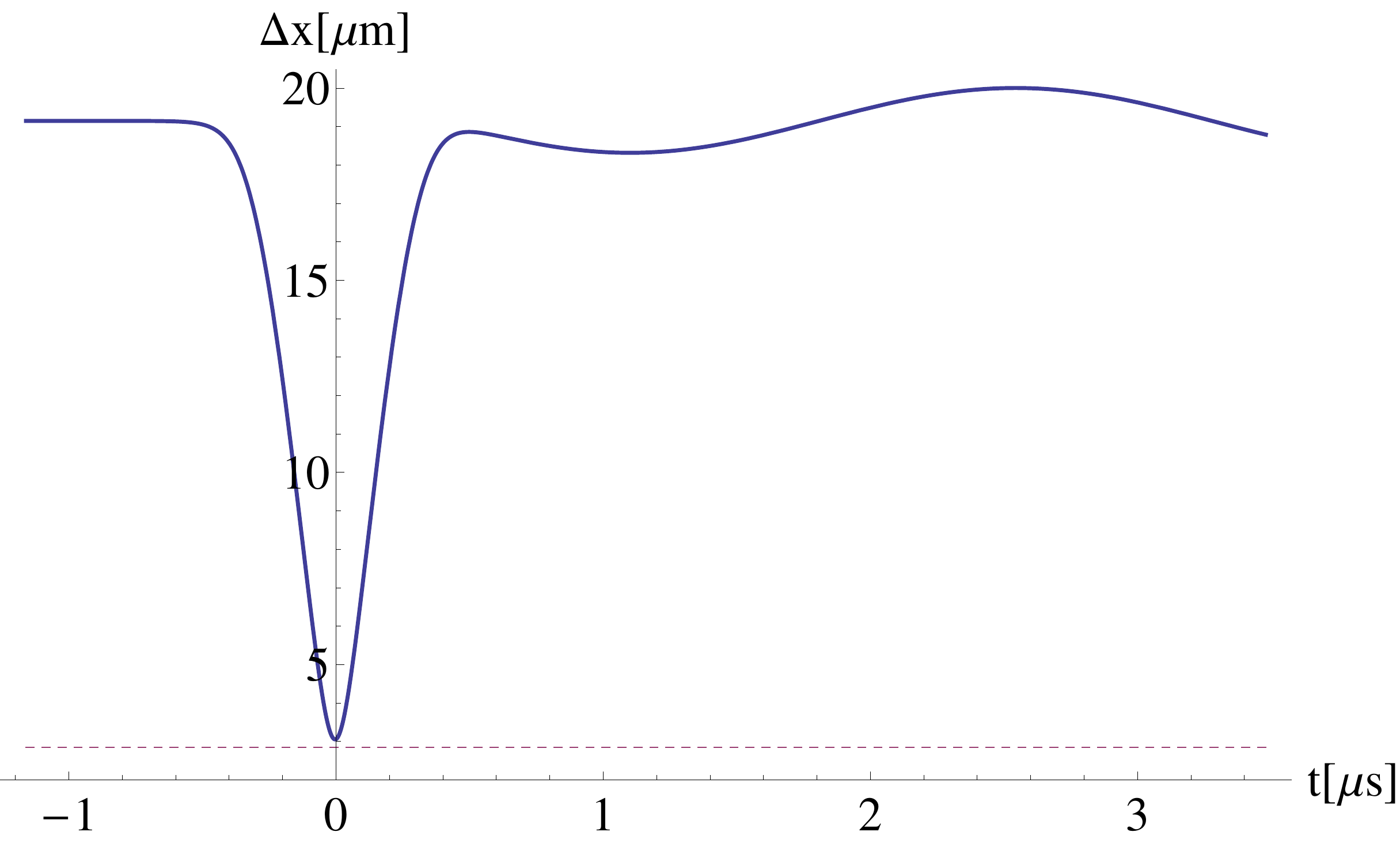}
\caption{Numerically calculated trajectory $\Delta x$ for two ions confined
in the axial potential of its characteristic $\omega_\text{ax}(t)$ presented in Fig.~\ref{fig:wax}.
After the collision the ions oscillate relative to their initial equilibrium
positions.
The red dashed line shows the critical distance at which $\Omega_-=0$ where the ion chain becomes instable.}
\label{fig:delta_x}
\end{figure}

\begin{figure}
\includegraphics[width=0.95\linewidth]{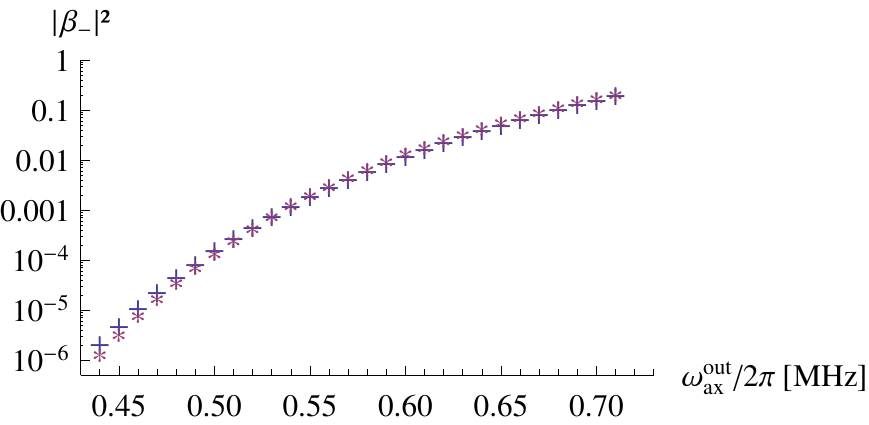}
\caption{Bogoliubov coefficient $\beta_-$ obtained (+) numerically or (*)
based on Eq. (\ref{eqn:betap1p2}) for collisions induced by an axial
potential as presented in Fig. \ref{fig:wax}, where the peak confinement
$\omega_ \text{ax}^\text{max}$ is varied.}
\label{fig:comparision}
\end{figure}

Let us exploit these results to propose a realistic implementation:
a collision of two ${}^{25}$Mg${}^+$ ions.
%
%
%
trapped in a radial potential with frequency
$\omega_\text{rad}=2\pi \cdot 3.5$ MHz and an initial axial potential with
frequency $\omega_\text{ax}^\text{in}=  2 \pi \cdot 0.2$ MHz.
The initial equilibrium distance is $\Delta x^\text{eq}\approx 19.1 ~\mu \text{m}$.
%
As an example we consider the axial confinement presented in Fig.
\ref{fig:wax}, where we increase $\omega_\text{ax}^\text{in}$ in
approximately 0.5 $\mu$s to $\omega_\text{ax}^\text{max}= 2 \pi
\cdot 0.7$ MHz, keep it constant for around 0.5 $\mu$s and return to
$\omega_\text{ax}^\text{in}$.
In Fig. \ref{fig:delta_x} the resulting ion trajectory is illustrated.
The exact Bogoliubov coefficient can be evaluated either numerically to
$|\beta_-|^2 \approx 0.18$ or approximately based on (\ref{eqn:betap1p2})
to $|\beta_-|^2 \approx 0.2$.
A more extensive comparison between the approximation
(\ref{eqn:betap1p2}) and the exact numerical results can be carried
out by calculating the Bogoliubov coefficients for different final
confinement $\omega_\text{ax}^\text{max}$ by both methods.
The result is illustrated in Fig. \ref{fig:comparision} and
indicates that we achieve good agreement over several orders of
magnitude.

Furthermore this realistic results permit predicting that the mean
phonon numbers created in the radial mode can be five times larger
than the residual thermal excitation of $n_\text{th}\approx 0.05$,
achievable by current cooling techniques. In addition, the
characteristic phonon distribution of the squeezed state allows to
clearly distinguish the pairwise created phonons from the thermal
background.
Therefore we conclude that analogue to cosmological particle
creation effects should be observable in already state-of-the-art
ion traps \cite{wineland_2012,schmidt-kaler_2012}.
%

So far we have only treated collisions of 2 ions.
However, collisions of the form (\ref{eqn:modb}) permit also exact analytical
expressions for the Bogoliubov coefficients of higher normal modes.
In the limit of slow and moderate collisions they can be approximated by
\begin{equation}
|\beta_\kappa|^2 \propto
\exp\left[-2 \pi \frac{\sqrt{\omega^2_\text{rad}-\omega^2_\kappa} - \displaystyle \Delta \Omega_\text{col} \frac{\omega_\kappa}{\omega_\text{ax}^\text{in}}}{\omega_\text{col}} \right] \ .
\end{equation}
Consequently particle creation in the $\kappa$-th normal mode becomes only
important if
\begin{equation}
\left(\sqrt{\omega^2_\text{rad}-\omega^2_\kappa} - \displaystyle \Delta \Omega_\text{col} \frac{\omega_\kappa}{\omega_\text{ax}^\text{in}}\right) = \mathcal{O}\left(\omega_\text{col}\right)  \ .
\label{eqn:krit_col_eta}
\end{equation}
For an increasing $\Delta \Omega_\text{col}$, considerable creation
of pairs of phonons occurs therefore firstly in the mode with the
highest $\omega_\kappa$.
For large $N$ in a linear chain of ions, this mode is called the
zig-zag mode.
\subsection{Expansion model}

Another type of axial motion, which corresponds to an expansion of
the mutual distance of the ions, is described by the scale function
\begin{equation}
b_\text{ex}(t)= \left(1- \frac{\Delta \Omega_\text{ex}^2}{2 \left(\omega_\text{ax}^\text{in}\right)^2} \big( \tanh\left(\omega_\text{ex} t\right)+1 \big)\right)^{-\frac{1}{3}} \ .
\label{eqn:bex}
\end{equation}
The parameter $\Delta \Omega_\text{ex}^2$ describes the induced jump
in the normal mode frequency $\Omega_-^2(t)$, whereas
$\omega_\text{ex}$ determines how fast the expansion evolves.
Inserting $b_\text{ex}(t)$ into (\ref{eqn:rockingmode}) yields a differential
equation that is discussed in \cite{birrell_quantum_1982} as an example for
cosmological particle creation.
It can be solved in terms of hypergeometric functions whose asymptotic
behaviour is known for $t \to \pm \infty$.
The Bogoliubov coefficient reads
\begin{equation}
|\beta^\text{ex}_-|^2= \frac{\sinh^2\left(\displaystyle\frac{\pi}{2} \frac{\Omega_\text{out}-\Omega_\text{in}}{\omega_\text{ex}}\right)}{\sinh\left(\pi \displaystyle \frac{\Omega_\text{in}}{\omega_\text{ex}}\right)\sinh\left(\pi \displaystyle \frac{\Omega_\text{out}}{\omega_\text{ex}}\right)} \ ,
\label{eqn:beta_ex}
\end{equation}
where
\begin{equation}
\Omega_\text{out}=\sqrt{\Omega^2_\text{in} + \Delta \Omega^2_\text{ex}} \ .
\end{equation}
E.g. for very large $\omega_\text{ex}$, that means for a sudden
quench, the Bogoliubov coefficient can be approximated to
\begin{equation}
|\beta_\text{ex}^2|\approx
\frac{(\Omega_\text{out}-\Omega_\text{in})^2}{4 \Omega_\text{in} \Omega_\text{out}}
\ ,
\label{eqn:beta_sprung_2}
\end{equation}
However, in the case of moderate and slow expansions, i.e.,
$\omega_\text{ex} \ll \Omega_\text{in}$ the Bogoliubov coefficients
become
\begin{equation}
|\beta_\text{ex}|^2 \propto e^{-2 \pi \Omega_\text{in}/\omega_\text{ex}} \ .
\end{equation}

\section{Ion-Ion entanglement}

After having discussed the excitation process of pairs of phonons in
the last section, we now analyze the conditions to reach
entanglement between the ions and how robust this entanglement is
against thermal disturbances.
%
%
We discuss exclusively the case of $N=2$ ions.

We consider a system, that is initially in a thermal state with sufficiently separated ions to consider them initially
uncoupled, i.e., $\Omega_-(t_\text{in}) = \omega_\text{rad}$.
In this case the operators $\hat \delta y_\pm$ and $\hat \delta y_{1/2}$ form
two equivalent sets of normal modes and their corresponding initial creation
and annihilation operators are linked via
\begin{equation}
\hat{a}^\text{in}_+
=
\frac{1}{\sqrt{2}} \left(\hat{a}^\text{in}_1 + \hat{a}^\text{in}_2 \right)
\label{eqn:ap_a12}
\end{equation}
and
\begin{equation}
\hat{a}^\text{in}_-
=
\frac{1}{\sqrt{2}} \left(\hat{a}^\text{in}_1 - \hat{a}^\text{in}_2 \right) \ .
\label{eqn:am_a12}
\end{equation}
Next the system becomes squeezed, for example by an ion collision as discussed
in section \ref{sec:excitation_models}.
Finally, the ions return to their initial positions, such that the
coupling vanishes again.

Firstly, we focus on a small squeezing parameter $\xi$ and small
thermal excitations on within the radial mode
\begin{equation}
n_\text{th}
=
\left<\hat{n}_1 + \hat{n}_2 \right>
=
2 \left<\hat{n}_1\right>=2 \left<\hat{n}_2\right>
=
2 \coth\left(\frac{\hbar \omega_\text{rad}}{2 k_B T}\right) \ .
\label{eqn:n_th}
\end{equation}
Here $T$ is the (initial) temperature and $k_B$ is the Boltzmann constant.
We do not consider effects of thermal excitations in the axial modes because there is no coupling between axial and radial normal modes, see Eq. (\ref{eqn:radial}).
The initial density operator can then be written as
\begin{alignat}{1}
\hat{\rho}_\text{th}^\text{in}
&=
\left(1-n_\text{th}\right) +\ket{0}_1\ket{0}_2\bra{0}_1\bra{0}_2
\nonumber \\
&+
\frac{n_\text{th}}{2} \ket{1}_1\ket{0}_2\bra{1}_1\bra{0}_2
+ \frac{n_\text{th}}{2} \ket{0}_1\ket{1}_2\bra{0}_1\bra{1}_2
\nonumber \\
&+ \mathcal{O}(n_\text{th}^2) \ .
\label{eqn:rho_therm_verschr}
\end{alignat}
After the squeezing process described by the operator
$\hat{\mathcal{S}}_{\vec{\xi}}$ in Eq.~\eqref{eqn:squeez_ionen_aktiv},
the final density operator reads
\begin{equation}
\hat{\rho}_\text{th}^\text{out}
=
\hat{\mathcal{S}}_{\vec{\xi}} \hat{\rho}_\text{th}^\text{in}
\hat{\mathcal{S}}_{\vec{\xi}}^\dagger \ .
\end{equation}
The partially transposed matrix of $\hat{\rho}_\text{th}^\text{out}$
possesses the eigenvalues $n_\text{th} \pm |\xi_-|$ and becomes consequently
negative definite for sufficiently large $|\xi_-|$.
With the Peres-Horodecki-criterion \cite{horodecki_separability_1996},
which is a sufficient separability criterion for Gaussian states
\cite{simon_peres-horodecki_2000}, it follows, that the ions are entangled
if and only if $|\xi_-| >  n_\text{th}$.

The former result was obtained by assuming small parameters $|\xi_-|$ and
$n_\text{th}$.
However for Gaussian states such as thermal states and squeezed
thermal states (which we consider in our scenario), it is also
possible to evaluate the Peres-Horodecki-criterion for finite
parameters.
This was demonstrated in
\cite{serafini_symplectic_2004,adesso_entanglement_2007} and recently applied
to analogue gravity experiments in \cite{bruschi_robustness_2013}.
In appendix \ref{sec:covariance_fromalism} we adapt the formalism to our
system and conclude that an initial thermal state with thermal excitations
$n_+$ and $n_-$ in the $\delta \hat q_\pm$ normal modes becomes entangled
during a squeezing process in the $\delta \hat q_-$ mode if and only if the
squeezing parameter satisfies
$\sqrt{1+2n_-} \sqrt{1+2n_+} \exp \left(-|\xi_-|\right) <1$.
As expected, in the limit of small squeezing parameters and small thermal
excitations $2n_+=2n_-=n_\text{th}$, this results coincides with the former
entanglement criterion  $|\xi_-| >  n_\text{th}$.


\section{Conclusions}

We considered the radial modes of two or more ions in a trap which
we accelerate in the axial direction.
The shaping of the axial motion permits us to control the
time-dependent coupling between the radial fluctuations and to
create an characteristic excitation in these modes.
An advantage of this set-up in comparison to previous proposals lies
in exploiting axial and radial motion to allow us to
derive realistic parameters to enable the detection of phonon pair
creation. This will permit to investigate physics and test proposed
schemes as well to create entanglement between ions.
The process of phonon pair creation has been predicted to emerge in
an analogous way in cosmological particle creation or black hole
evaportaion, where the entanglement between the partners is related
to the entropy of the black hole.




\begin{thebibliography}{1}
\bibitem{birrell_quantum_1982}
N.~D. Birrell, P.~C.~W. Davies, {\em Quantum fields in curved space}
(Cambridge University Press, Cambridge, England, 1982).

\bibitem{reznik_2005}
A.~Retzker, J.~I.~Cirac, B.~Reznik, {\em Phys. Rev. Lett.} {\bf 94}, 050504
 (2005).

\bibitem{schutzhold_analogue_2007}
R.~Sch\"utzhold, \textit{et al.}, {\em Phys. Rev. Lett.} {\bf 99}, 201301
 (2007).

\bibitem{milburn_2005}
P.~M.~Alsing, J. P. Dowling,  G. J. Milburn, {\em Phys. Rev. Lett.} {\bf 94}, 220401
 (2005).

\bibitem{schaetz_2007}
L. Lamata, J. Le\'on, T. Sch\"atz, and E. Solano, {\em Phys. Rev. Lett.} {\bf 98}, 253005
 (2007).

 \bibitem{solano_2011}
J. Casanova {\textit et al.}, {\em Phys. Rev. Lett.} {\bf 107}, 260501
 (2011).

 \bibitem{walther_1992}
G.~Birkl, S.~Kassner and H.~Walther, {\em Nature (London)} {\bf 357}, 310
 (1992).
 
 
 
\bibitem{schmidt-kaler_2012}
A.~Walther, \textit{et al.}, {\em Phys. Rev. Lett.} {\bf 109}, 080501
 (2012).
 
\bibitem{wineland_2012}
R.~Bowler, \textit{et al.}, {\em Phys. Rev. Lett.} {\bf 109}, 080502
 (2012). 
 

\bibitem{horodecki_separability_1996}
M.~Horodecki, P.~Horodecki, R.~Horodecki,
{\em Phys. Lett. A} {\bf 223}, 1 (1996).

\bibitem{simon_peres-horodecki_2000}
R.~Simon, {\em Phys. Rev. Lett.} {\bf 84}, 2726 (2000).

\bibitem{adesso_entanglement_2007}
G.~Adesso, F.~Illuminati,
{\em J. Phys. A: Math. Theor.} {\bf 40}, 7821 (2007).

\bibitem{serafini_symplectic_2004}
A.~Serafini, F.~Illuminati, S.~De~Siena,
{\em J. Phys. B: At. Mol. Opt. Phys.} {\bf 37}, L21 (2004).

\bibitem{bruschi_robustness_2013}
D.~E. Bruschi, N.~Friis, I.~Fuentes, S.~Weinfurtner,
{\em New J. Phys.} {\bf 15}, 113016 (2013).

\bibitem{schuetzhold_2013}
R. Sch\"utzhold, W.~G. Unruh, in: {\em Analogue Gravity Phenomenology},
Lecture Notes in Physics {\bf  870}, 51 (2013).

\end{thebibliography}


\appendix
\section{Bogoliubov coefficients}
\label{sec:beta_col}

The solutions of the rocking mode differential equation
(\ref{eqn:rockingmode}) for the collision model (\ref{eqn:modb}) are the
associated Legendre polynomials $P_\mu^\nu(z)$ with the substitutions
\bea
z&=&\tanh(t)
\,,
\nn
\mu &=& i \frac{\Omega_\text{in}}{\omega}
\,,
\nn
\nu &=&\frac{1}{2}
\left(i \sqrt{4 \Delta\Omega_\text{col}^2 / \omega_\text{col}^2-1} - 1\right)
\,.
\ea
Their asymptotic behavior is
\begin{equation}
P_\mu^\nu\big(\tanh(t)\big)
\overset{t \to \infty}{\longrightarrow}
\frac{e^{\mu t}}{\Gamma(1-\mu)}
\end{equation}
and
\begin{xalignat}{1}
P_\mu^\nu\big(\tanh(t)\big)
\overset{t \to -\infty}{\longrightarrow}
&
\frac{\Gamma(-\mu)}{\Gamma(-\mu -\nu) \Gamma(1- \mu +\nu)}e^{\mu t}
\nonumber  \\
&- \frac{\sin(\pi \nu)\Gamma(\mu)}{\pi} e^{-\mu t} \ .
\end{xalignat}
The Bogoliubov coefficient $\beta_-^\text{col}$ is therefore
\begin{equation}
\beta_-^\text{col}
=
- \frac{ \sin(\pi \nu) \Gamma(\mu)}{\pi \Gamma(1-\mu)}
=
\frac{\sin(\pi \nu)}{\sin(\pi \mu)} \ .
\end{equation}
Back substitution yields finally
\begin{equation}
|\beta_-^\text{col}|^2
=
\left|
\frac{\cosh \bigg(\displaystyle\frac{\pi}{2} \sqrt{\frac{4 \Delta \Omega_\text{col} ^2}{\omega_\text{col} ^2}-1}\bigg)}{\sinh \left(\displaystyle\frac{\pi \Omega_\text{in}}{\omega_\text{col} }\right)}
\right|^2 \ .
\end{equation}

\section{WKB-approximation}
\label{sec:WKB}

We derive here the general exponential behavior of the Bogoliubov
coefficients for slow and moderate collisions in a normal mode with
frequency $\Omega^2(t)$.
Moderate means that we stay away from the critical point, i.e.,
\begin{equation}
\Omega^2(t) > 0 \ ,
\label{eqn:wkb_assump1}
\end{equation}
while slow means that
\begin{equation}
\left| \frac{\dot{\Omega}(t)}{\Omega^2(t)} \right|  \ll 1 \ .
\label{eqn:wkb_assump2}
\end{equation}
For a typical collision $\Omega^2(t)$ reaches its minimum when the ions are
closest and the scale functions becomes minimal.
Without loss of generality this happens at $t=0$.
As shown in \cite{schuetzhold_2013}, under these conditions a
WKB-approximation yields the exponential behavior of the Bogoliubov
coefficient as
\begin{equation}
|\beta|^2
\propto
\exp\left[-4 \Im\left\{\displaystyle\int_0^{t_*} \Omega(t) dt \right\} \right]
\ ,
\label{eqn:WKB_appendix}
\end{equation}
where $t_*$ denotes the root of $\Omega(t)$ in the upper complex plane.
Phonon creation happens more likely when the exponent is small.
This can be achieved by working with low frequencies $\Omega(t)$ and small
values for $t_*$.

Next, we calculate the exponent explicitly for collisions that are well
described by a Taylor expansion
\begin{equation}
\Omega^2(t)\approx \Omega^2_\text{min} + \frac{1}{2}K^2 t^2 \ ,
\label{eqn:WKB_minimumentwicklung}
\end{equation}
in the region $|t|<|t_*|$, where
\begin{equation}
K^2=\frac{d^2}{dt^2}\Omega^2(t)\Big|_{t=0}
\end{equation}
is the curvature.
Their complex root is approximated by
\begin{equation}
t_* \approx i \sqrt{2} \, \frac{\Omega_\text{min}}{K} \ .
\label{eqn:turningpoint}
\end{equation}
Finally, evaluating (\ref{eqn:WKB_appendix}) leads to
\begin{equation}
|\beta|^2 \propto
\exp\left[-\sqrt{2} \pi \frac{\Omega^2_\text{min}}{K} \right] \ .
\label{eqn:wkb_approx_end}
\end{equation}
For the model collision with $b_\text{col}(t)$ in (\ref{eqn:modb}),
this yields the exponential behavior
\begin{equation}
|\beta^\text{col}_-|^2 \propto
\exp\left[
- 2 \pi \frac{(\Omega_\text{in}-\Delta \Omega_\text{col})}{\omega_\text{col}}
\right] \ ,
\end{equation}
in agreement with (\ref{eqn:exponent_kriterium_col}).

\section{Covariance matrix formalism}
\label{sec:covariance_fromalism}

To apply the entanglement criteria for Gaussian states developed in
\cite{adesso_entanglement_2007,serafini_symplectic_2004} to our system we
define the phase space vector with respect to the ion coordinates
\begin{equation}
\vec{\hat{R}}_{12}
=
\left(
\begin{matrix}
\delta\hat{y}_1 &\delta\dot{\hat{y}}_1 &\delta\hat{y}_2 &\delta\dot{\hat{y}}_2
\end{matrix} \right)^T \ .
\end{equation}
The corresponding covariance matrix reads
\begin{equation}
\sigma_{kl}:= \frac{1}{2} \left< \hat{R}_k \hat{R}_l +\hat{R}_l \hat{R}_k\right>
\,.
\label{eqn:kov_matrix}
\end{equation}
We also define the phase space vector with respect to the normal coordinates
\begin{equation}
\vec{\hat{R}}_\pm
=
\left(\begin{matrix}
\delta \hat{y}_+ &\delta \hat{p}_+ &\delta \hat{y}_- &\delta \hat{p}_-
\end{matrix} \right)^T
=
\matr{D}\cdot\hat{\vec{R}}_{12} \ ,
\end{equation}
with the transformation matrix
\begin{equation}
\matr{D}
=
\frac{1}{\sqrt{2}}
\left(
\begin{matrix}1&0&1&0\\0&1&0&1\\1&0&-1&0\\1&0&-1&0 \end{matrix}
\right) \ .
\end{equation}
The covariance matrices corresponding either to $\vec{\hat{R}}_{12}$ or to
$\vec{\hat{R}}_\pm$ are linked via
\begin{equation}
\matr{\sigma}_{12} = \matr{D}\cdot\matr{\sigma}_\pm\cdot\matr{D}
\label{eqn:sigma12pm} \ .
\end{equation}
We consider an initial thermal covariance matrix
\begin{equation}
\matr{\sigma}_{\pm}^\text{in}= \frac{1}{2}
\left(\begin{matrix}
1+2n_+ &0&0&0\\
0&1+2n_+ &0&0\\
0&0&1+2n_- &0\\
0&0&0&1+2n_-\\
\end{matrix}\right) \ .
\end{equation}
with the thermal occupation numbers
\begin{equation}
n_\pm =\coth\left(\frac{\hbar \Omega_{\text{rad}\pm}}{2 k_B T} \right) \ .
\end{equation}
Its time evolution during a squeezing process is
\begin{equation}
\matr{\sigma}_\pm^\text{out}
=
\matr{S}_\pm\cdot\matr{\sigma}_\pm^\text{in}\cdot\matr{S}_\pm^T \ .
\end{equation}
where $\matr{S}_\pm$ is a symplectic matrix containing the Bogoliubov
coefficients
\begin{equation}
\matr{S}_\pm
=
\left(\begin{matrix}
\phantom{-}\Re\{\alpha_+\} &\Im\{\alpha_+\}&0&0
\\
-\Im\{\alpha_+\} &\Re\{\alpha_+\}&0&0
\\
0&0&\phantom{-}\Re\left\{\alpha_- - \beta_- \right\} &\Im\left\{\alpha_- + \beta_- \right\}
\\
0&0&-\Im\left\{\alpha_- - \beta_- \right\} &\Re\left\{\alpha_- + \beta_- \right\}\\
\end{matrix}\right) \ .
\end{equation}
Hence we get
\begin{equation}
\matr{\sigma}_{12}^\text{out}
=
\matr{D}\cdot\matr{S}_\pm\cdot\matr{D}\cdot\matr{\sigma}_{12}^\text{in}
\cdot\matr{D}\cdot \matr{S}_\pm^T\cdot\matr{D} \ .
\end{equation}
As shown in \cite{adesso_entanglement_2007}, for Gaussian states the
Peres-Horodecki criterion can be formulated as a criterion on the two
symplectic eigenvalues $\lambda_\pm$ of the partial transposed convariance
matrix
\begin{equation}
\left(\matr{\sigma}_{12}^\text{out}\right)^{PT}
=
\matr{T}\cdot\matr{\sigma}_{12}^\text{out}\cdot\matr{T}
\end{equation}
with $\matr{T}=\text{diag}({1,-1,1,1})$.
The ions are entangled if one of the symplectic eigenvalues is smaller
than $1/2$.

For our system we obtain the symplectic eigenvalues as the two positive
eigenvalues of $i\matr{J}\cdot\left(\matr{\sigma}_{12}^\text{out}\right)^{PT}$
to
\begin{alignat}{1}
\lambda^{PT}_{\pm}
&=
\frac{1}{2} \sqrt{1+2 n_-}\sqrt{1+2 n_+} \left(|\alpha_-| \pm |\beta_-|\right)^2
\nonumber \\
&=
\frac{1}{2} \sqrt{1+2 n_-}\sqrt{1+2 n_+} \ e^{\pm |\xi_-|} \ ,
\end{alignat}
where
\begin{equation}
\matr{J}= \left(
\begin{matrix}0&0&1&0\\0&0&0&1\\-1&0&0&0\\0&-1&0&0 \end{matrix}
\right) \ .
\end{equation}
Therefore the ions are entangled if
$\sqrt{1+2n_-} \sqrt{1+2n_+} \exp \left(-|\xi_-|\right) <1$.

Furthermore, in the case of symmetric squeezing, the Entanglement of
Formation $E_F$ can be evaluated explicitly \cite{serafini_symplectic_2004}.
Squeezing is called symmetric when the two blockdiagonal $2\times 2$ matrices
of $\matr{\sigma}_{12}^\text{out}$ posses identical determinants, which is here
the case.
The Entanglement of Formation is then
\begin{equation}
E_F= \begin{cases}
f(\lambda_-^{PT})&\text{wenn }0 < \lambda^{PT}_- <\frac{1}{2}\\
0&\text{wenn } \frac{1}{2} \le \lambda_-^{PT} \ ,
\end{cases}
\end{equation}
with the function
\begin{equation}
f(x)
=
\frac{\left(\frac{1}{2}+x\right)^2}{2x}
\ln\left(\frac{\left(\frac{1}{2}+x\right)^2}{2x}\right)
-\frac{\left(\frac{1}{2}-x\right)^2}{2x}
\ln{\left(\frac{\left(\frac{1}{2}-x\right)^2}{2x} \right)} \ .
\label{eqn:entaglement_of_formation}
 \end{equation}

\end{document}